\newif\iffigs\figstrue
\documentclass[11pt,a4paper]{article}
\usepackage{amsmath}
\usepackage{amsthm}
\usepackage{amssymb}
\usepackage{graphicx}

\usepackage{epstopdf}
\begin{document}

\newcommand{\beq}[1]{\begin{equation}\label{#1}}
\newcommand{\eq}{\end{equation}}

\newtheorem{predl}{Proposition}
\newtheorem{lem}{Lemma}[section]
\newtheorem{theor}{Theorem}[section]

\begin{center}

\vspace{20mm}

{\Large{\sc Integrable Cosmological Potentials}
}\\


\vspace{35pt}
{\sc V. V. Sokolov$^a$ and A. S. Sorin$^{b,c,d}$}\\[15pt]

{${}^a$\small L.D. Landau Institute for Theoretical Physics, \\
142432 Chernogolovka (Moscow region), Russia\\}e-mail: {\small \it
vsokolov@landau.ac.ru}\vspace{10pt}

{${}^b$\small Bogoliubov Laboratory of Theoretical Physics and\\
Veksler and Baldin Laboratory of High Energy Physics\\
Joint Institute for Nuclear Research \\ 141980 Dubna, Moscow Region, Russia
\\ }e-mail: {\small \it
sorin@theor.jinr.ru}\vspace{10pt}

{${}^c$\small National Research Nuclear University MEPhI\\
(Moscow Engineering Physics Institute),\\
Kashirskoe Shosse 31, 115409 Moscow, Russia}\vspace{10pt}

{${}^d$\small Dubna International University, \\
141980 Dubna (Moscow region), Russia}\vspace{10pt}

\vspace{3mm}

\end{center}

\begin{abstract}
The problem of classification of the Einstein--Friedman cosmological Hamiltonians $H$ with a single scalar inflaton field $\varphi$ that possess an additional integral of motion polynomial in momenta on the shell of the Friedman constraint $H=0$ is considered. Necessary and sufficient conditions for the existence of first, second, and third degree integrals are derived. These conditions have the form of ODEs for the cosmological potential $V(\varphi)$. In the case of linear and quadratic integrals we find general solutions of the ODEs and construct the corresponding integrals explicitly. A new wide class of Hamiltonians that possess a cubic integral is derived. The corresponding potentials are represented in a parametric form in terms of the associated Legendre functions. Six families of special elementary solutions are described and sporadic superintegrable cases are discussed.
\end{abstract}


\newpage
\tableofcontents
\newpage

\section{Introduction}

The present paper is devoted to integrable cases for the Einstein equation with the Friedmann--Lemaitre--Robertson--Walker  metric and  a scalar field.  The 4-dimensional Einstein equation with a scalar inflaton field $\Phi(x)$ can be derived from the Lagrangian
\begin{equation}\label{GenL}
 L\,=\,\sqrt{- \, \mbox{det} \,g} \left( \frac{1}{2}\,R \, + \, \frac{1}{2}\, g^{\mu\nu} \, \partial_\mu \,\Phi \ \partial_\nu \, \Phi \,
   - \, \frac{2}{3}\,\mathcal{V}(\Phi) \right),
\end{equation}
where  $g^{\mu\nu}(x)$ is a metric,  $R_{\mu \nu}$ and $R$ are its Ricci curvature tensor and scalar curvature, respectively.
The variation of the Lagrangian with respect to $g^{\mu\nu}(x)$ and $\Phi$ leads to the Einstein equations
\begin{eqnarray}
&& R_{\mu \nu}\,-\,\frac{1}{2}\,R\,g_{\mu \nu}\,=\,\,\partial_{\mu}\Phi\,\partial_{\nu}\Phi\,-\,
\Big(\frac{1}{2}\,g^{\rho \sigma}\partial_{\rho}\Phi\,
\partial_{\sigma}\Phi \,-\,\frac{2}{3}\,\mathcal{V}(\Phi)\Big)\,g_{\mu \nu},\nonumber\\
&& \frac{1}{\sqrt{- \, \mbox{det} \,g}}\,\partial_{\nu}\Big(\sqrt{- \, \mbox{det} \,g}\,g^{\mu \nu}\partial_{\mu}\Phi\Big)\,+\,
\frac{2}{3}\,\frac{d}{d \Phi}\mathcal{V}(\Phi)=0.
\label{EinsteinKleinGordon}
\end{eqnarray}

For the cosmological applications (see books \cite{CosmoBooks}) the isotropic, homogeneous and spatially-flat  Friedmann--Lemaitre--Robertson--Walker  metric
\begin{equation}
d\,s^2\,=\,e^{\,2\,\mathcal{B}(t)}\, d\,t^2\,-\,e^{\frac{2}{3} \mathcal{A}(t)}\, d\vec{\mathbf{x}}^{~2}
\label{FLRW}
\end{equation}
is of a great importance. Here ${\cal B}(t)$ is a gauge function, which can be fixed by a reparametrization of time $t.$
For metric (\ref{FLRW}) and $\Phi=\sqrt{\frac{2}{3}}\,\varphi(t)\,$ the Lagrangian (\ref{GenL}) reduces to
\begin{eqnarray}
{\cal L}\,=\frac{2}{3}\,e^{\,{\cal A}\,-\,{\cal B} }\left(-\,\frac{1}{2}\,\dot{\cal A}^{\,2}\,+\, \frac{1}{2}\,\dot{\varphi}^2\,-\,e^{\,2\,{\cal B}}\,\mathcal{V}(\varphi) \right)\, .
\label{LagrS}
\end{eqnarray}
One can fix the gauge $\mathcal{B}\,=\,\mathcal{A}$, which we will use in what follows. Under this gauge
equations (\ref{EinsteinKleinGordon}) have the form
\begin{eqnarray}
&& \ddot{\varphi}\,+\,e^{\,2 {\cal A}}\,\mathcal{V}^{\,\prime}(\varphi)\,=\,0, \nonumber \\
&& \ddot{\cal A}\,-\,2\, e^{\,2 {\cal A}}\, \mathcal{V}(\varphi)\,=\,0, \nonumber\\
&& -\,\frac{1}{2}\,\dot{\cal A}^{\,2}\,+\, \frac{1}{2}\,\dot{\varphi}^{\,2}\,+\,e^{\,2{\cal A}}\,\mathcal{V}(\varphi)\,=\,0, \label{hamS}
\end{eqnarray}
where the signs $^{.}$ and $^{'}$ denote the derivatives $\frac{d}{dt}$ and $\frac{d}{d \varphi}$, respectively.
The latter equation is known as the Friedmann or the Hamiltonian constraint.

The Einstein--Friedmann cosmological models (\ref{LagrS}), (\ref{hamS}) possess effectively two physical degrees of freedom $\{\,Q_1\,=\,\mathcal{A}(t),\,Q_2\,=\, \varphi(t)\,\}.$ Equations  (\ref{hamS}) can be characterized by the Hamiltonian dynamics with the conjugate momenta $\{\,P_1\,=\, -\,\dot {\mathcal{A}} (t),\, P_2\,=\,\dot {\varphi}(t)\,\}$ and the Hamiltonian \begin{equation}\label {ham}
\mathcal{H}\,\equiv\, -\,\frac12 \,P_1^2 \,+\, \frac12 \, P_2^2\,+\,e^{2 Q_1}\, \mathcal{V}(Q_2)
\end{equation}
constrained to zero-level shell  $\mathcal{H}=0$ (see formula (\ref{hamS})).

Pioneering  inflationary models \cite{Starob,infl} are just based on the Einstein--Friedmann equations. A vast activity of investigations has produced a series of potentials $V(\varphi)$ (see, e.g. recent review \cite{Martin:2013tda}) that display phenomenologically attractive features. However explicit analytic solutions of the Einstein--Friedmann equations with such potentials are extremely scarce in the literature. Usually, an exact analytic solution of an integrable dynamical system, which approximate the original one, can often provide deeper insights in this system and a real understanding of fundamental physical phenomena encoded there can be attained in this way.

With such motivation in mind, in the present paper we perform a systematic search for potentials $\mathcal{V}(\varphi)$ such that the corresponding Hamiltonian (\ref{ham}) possesses an additional integral polynomial in momenta. Hereafter such potentials and   Hamiltonians are called the {\it integrable} ones.
For integrable potentials one can apply powerful methods of the theory of Liouville integrable dynamical systems  to derive explicit solutions for the corresponding Einstein--Friedmann equations (\ref{hamS}).

Applying the canonical transformation
$\{\,Q_1,\,Q_2,\,P_1,\,P_2\,\}\,\rightarrow \,
\{\,\frac{1}{2}\, q_1,\,\frac{1}{2}\, q_2,\, 2\, p_1,\, 2\, p_2\,\}$ and substituting  $\{\,\mathcal{H},\,\mathcal{V}(Q_2)\,\} \rightarrow \{\,-\,2\,H,\,-\,2\, V(q_2)\,\},$
we bring Hamiltonian   (\ref{ham}) to the form
\begin{equation}\label {HHour}
H\,\equiv\,  \,p_1^2 \,-\, p_2^2\,+\,e^{q_1}\, V(q_2).
\end{equation}

The classification problem for two-dimensional Hamiltonians of the form
\begin{equation} \label{HH}
H\,=\,p_1^2\,-\,p_2^2\,+\, U(q_1,q_2)
\end{equation}
that possess an additional integral of motion at the zero-level $H=0$ polynomial in momenta was considered by many authors (see, reviews \cite{hietarinta,perelomov} and references therein).
There exists a wide class of the following canonical transformations
\begin{eqnarray} \label{canon}
&& \tilde q_1\,=\,\alpha(q_1\,+\,q_2)\,+\,\beta(q_1\,-\,q_2), \qquad  \tilde q_2\,=\,\alpha(q_1\,+\,q_2)\,-\,\beta(q_1\,-\,q_2),\nonumber\\[3mm]
&&\tilde p_1\,=\, \frac{p_1\,(\alpha'\,+\,\beta')\,-\,p_2\, (\alpha'\,-\,\beta')}{4 \alpha' \beta'}, \quad \tilde p_2\,=\,\frac{p_2\,(\alpha'\,+\,\beta')\,-\,p_1\, (\alpha'\,-\,\beta')}{4 \alpha' \beta'},
\end{eqnarray}
which preserve the form of Hamiltonian (\ref{HH}) and the constraint $H=0$. Here $\alpha(q_1\,+\,q_2)$ and $\beta(q_1\,-\,q_2)$ are arbitrary functions of their arguments and the sign $'$ denotes the corresponding derivatives. Indeed, after such a transformation the Hamiltonian becomes
$$
H\,=\,\frac{1}{4 \alpha' \beta'}(\tilde p_1^2\,-\,\tilde p_2^2)\,+\, U(\tilde q_1,\tilde q_2).
$$
Taking into account the zero-level constraint $H=0,$ one can multiply $H$ by $4 \alpha' \beta'$ to bring $H$ to the original form (\ref{HH}) with the transformed potential $4 \alpha'\, \beta' \,\tilde U(\tilde q_1,\tilde q_2)$. Usually such canonical transformations are used to reduce coefficients at the highest powers of additional integral to constants. After that, the integrability conditions are presented in the form of a PDE for $U(q_1,q_2).$ Some sporadic explicit solutions of these PDEs are known.

Notice that for a given integrable Hamiltonian (\ref{HH}) it is a difficult task to verify whether or not there exists a transformation (\ref{canon}) that brings $H$ to the form (\ref{HHour}) and to derive this transformation explicitly. Nevertheless, in this way a number of integrable Hamiltonians (\ref{HHour}) were constructed in \cite{FSS}.

For the particular case (\ref{HHour}) we are interested in this paper the potential has a very special product structure $e^{q_1}\, V(q_2)$, which is form-invariant with respect to a very restricted subset of transformations (\ref{canon}). Namely, only the following admissible  transformations:
\begin{equation}\label{tr1}
q_1\to q_1+\tau_1, \qquad q_2\to q_2+\tau_2
\end{equation}
and
 \begin{equation}\label{tr2}
 q_2\to - q_2, \qquad p_2\to - p_2,
 \end{equation}
where $\tau_1$ and $\tau_2$ are arbitrary constants,  preserve the structure of the potential. Transformation (\ref{tr1}) corresponds to the functions $\{\alpha(x)\,=\,\frac{1}{2}(x\,+\,\tau_1\,+\,\tau_2),\,
\beta(x)\,=\,\frac{1}{2}(x\,+\,\tau_1\,-\,\tau_2)\}$ in (\ref{canon}).

The paper is organized as follows. In Section 2 we consider Hamiltonians of the special form (\ref{HHour}) that possess an additional integral of first degree in momenta at the level $H=0$. We prove that this fact is equivalent to the potential $V(q_2)$ to be a solution of a linear master ODE of first order. Coefficients of the master equation depend on two parameters. One of these parameters is continuous and another  projective parameter can be reduced to 0 or to $\pm 1$ by a scaling.
We find solutions of the master equation corresponding to all possible values of the parameters. A comparison of the results shows that integrable potentials from Section 2  were found in \cite{FSS} in a different way.

Section 3 is devoted to Hamiltonians (\ref{HHour}) with integrals of second degree. In this case the master equation turns out to be a linear ODE of second order. We find all its solutions and prove that there are no other integrable cases besides ones  presented in \cite{FSS,FS2}.

In Section 4.1 we construct a master equation for the case of cubic integrals. In contrast with the previous cases, it is a quite complicated nonlinear equation, which contains two parameters just as in Sections 2 and 3.  In Section 4.2, when the projective parameter is equal to zero, we reduce the master equation to a linear ODE of  second order and solve it in terms of the associated Legendre functions. We also find particular solutions in elementary functions and new integrable potentials corresponding to them.

The simplest example of the  Hamiltonians found in Section 4.2 has the form
\begin{equation}\label{new}
H\,=\,p_1^2\,-\,p_2^2\,+\, c_1\, e^{q_{1}}\, \Big( e^{-\frac{q_{2}}{12}}\,-\,c_2 e^{\frac{3 \,q_{2}}{4}}  \Big)^2,
\end{equation}
where $c_1,\,c_2$ are arbitrary constants. For $c_1>0$ the potential is positive definite.
Such potentials are important in the context of the cosmological model building in supergravity \cite{FKLP}. For example, the famous Starobinsky potential \cite{Starob}\footnote{This potential occurs through the dual-equivalent representation \cite{WB} of $R\,+\,\gamma\,R^2$ Starobinsky original model \cite{Starob,KLSM} in terms of the usual Einstein gravity and a scalar field.}
\begin{equation}\label{StarPotEquiv}
V(q_2)\,=\,c_1\,\Big(1\,-c_2\,e^{-{\frac{q_2}{3}}}  \Big)^2 \nonumber
\end{equation}
belongs to this class.
It follows from the results of our paper that the Starobinsky Hamiltonian does not possess additional linear, quadratic and cubic integrals.

In Section 5 we present plots of several interesting integrable potentials found in Section 4. Moreover, we extract
three superintegrable Hamiltonians from the ones obtained in Sections 2,3, and 4.

\section{Hamiltonians with linear integrals}

\subsection{Master equation and its general solution}

Suppose that Hamiltonian (\ref{HH}) at the level $H=0$ admits an additional integral of motion of the form
\begin{equation}\label{lin}
I_1=A(q_1,q_2)\,p_1+B(q_1,q_2)\,p_2+C(q_1,q_2).
\end{equation}
Then the condition $\{H,I_1\}=0\, \,({\rm mod}\, H=0)$ is equivalent to the following relations:
\begin{equation}\label{con1}
\frac{\partial A}{\partial q_1}=\frac{\partial B}{\partial q_2}, \qquad \frac{\partial A}{\partial q_2}=\frac{\partial B}{\partial q_1},
\qquad \frac{\partial C}{\partial q_1}=\frac{\partial C}{\partial q_2}=0,
\end{equation}
\begin{equation}\label{con2}
A \frac{\partial U}{\partial q_1}+B \frac{\partial U}{\partial q_2}+2 \frac{\partial A}{\partial q_1}\,U=0.
\end{equation}
Solving (\ref{con1}), we find that
\begin{equation}\label{sr}
A(q_1,q_2)=r(q_1+q_2)+s(q_1-q_2), \qquad B(q_1,q_2)=r(q_1+q_2)-s(q_1-q_2),
\end{equation}
and  $C(q_1,q_2)=const$.

{\bf Proposition 1.} The potential $V(q_2)$ obeys the following ODE:
\begin{equation}\label{eqlin}
V'+(2\mu+1)\, \frac{C(q_2)}{S(q_2)}\,\,V=0,
\end{equation}
where
\begin{equation}\label{ch}
C(q_2)\,\equiv\,\frac{1}{2}\,(a\, e^{\mu \, q_2}\,+\,b \,e^{-\mu \, q_2}), \qquad S(q_2)\,\equiv\,\frac{1}{2}\,(a\, e^{\mu \, q_2}\,-\,b \,e^{-\mu \, q_2}),
\end{equation}
$a,\,b,$ and $\mu$ are constants.

{\bf Proof.} For $U=e^{q_{1}}\, V(q_2)$ equation (\ref{con2}) reduces to
\begin{equation}\label{maseq1}
B\, V'(q_2)+A\, V(q_2)+2 \frac{\partial A}{\partial q_1} V(q_2)=0.
\end{equation}
Differentiating this relation by $q_2$ and using (\ref{con1}), we get
\begin{equation}\label{funeq}
2 V\, \frac{\partial^2 B}{\partial q_1^2}+V\,\frac{\partial B}{\partial q_1}+V''\,B+3 V'\,\frac{\partial A}{\partial q_1}+V'\,A=0.
\end{equation}
For fixed $q_2$ these two relations form a system of ODEs with constant coefficients  for $A$ and $B$. Therefore both $A(q_1,q_2)$ and $B(q_1,q_2)$
are quasi-polynomials in $q_1$. It follows from (\ref{sr}) that the functions $r(q_1+q_2)$ and $s(q_1-q_2)$ are quasi-polynomials and
\begin{equation}\label{AAA}
A(q_1,q_2)\,=\,\sum R_j(q_1+q_2) e^{\lambda_j (q1+q2)} +\sum T_j(q_1-q_2) e^{\lambda_j (q1-q2)},
\end{equation}
$$
B(q_1,q_2)\,=\,\sum R_j(q_1+q_2) e^{\lambda_j (q1+q2)} -\sum T_j(q_1-q_2) e^{\lambda_j (q1-q2)},
$$
where $R_j(q_1+q_2)\,=\,a_j \,(q_1+q_2)^{m_{j}}+\cdots,\quad T_j(q_1-q_2)\,=\,b_j\, (q_1-q_2)^{m_{j}}+\cdots$ are polynomials. Substituting these expressions for $A(q_1,q_2)$ and $B(q_1,q_2)$ into (\ref{maseq1}) and collecting terms at $q_1^{m_{j}} e^{\lambda_j q_1} $, we get (\ref{eqlin}) with $\mu=\lambda_j, a=a_j, b=b_j.$ $\square$

Equation (\ref{eqlin}) is also a sufficient condition for the integrability of Hamiltonian (\ref{HHour}). It can be easily verified that if a potential $V$ satisfies (\ref{eqlin}), then for
\begin{equation}\label{rs}
r(q_1+q_2)=a\,e^{\mu (q_1+q_2)}, \qquad s(q_1-q_2)=b\,e^{\mu (q_1-q_2)}
\end{equation}
relations (\ref{con1}), (\ref{con2}) are fulfilled and Hamiltonian (\ref{HHour}) possesses the integral (\ref{lin}). Notice that involution (\ref{tr2}) in (\ref{eqlin}) corresponds to $a \leftrightarrow b.$ If equations (\ref{sr}), (\ref{maseq1}) possess other solutions besides (\ref{rs}), then Hamiltonian (\ref{HHour}) might be superintegrable.

In the generic case we can normalize $b=a$ or $b=-a$ by a shift (\ref{tr1}) of $q_2$. In the first and second case we get
\begin{equation}\label{deg1}
V(q_2)=c_1 \, (sh \,\mu q_2)^{-\frac{2\, \mu+1}{\mu}}, \qquad V(q_2)=c_1 \, (ch\,\mu q_2)^{-\frac{2 \,\mu+1}{\mu}},
\end{equation}
respectively.
The constant $c_1$ can be normalized to $\pm 1$ by (\ref{tr1}).

Degenerations $\mu=0$, $\mu=-1/2$ and $a b=0$ lead to exponential potentials
\begin{equation}\label{equ}
V(q_2) \,=\, c_1 e^{\lambda \,q_2},
\end{equation}
where $\lambda$ and $c_1$ are constants.

\subsection{Exponential potentials: maximal superintegrability and dynamical symmetry}

Consider potential (\ref{equ}) with $c_1=1.$ Let us show that the corresponding Hamiltonian $H$ is maximally superintegrable at the level $H=0$. To find all integrals of first degree, we  solve the linear functional equation  (\ref{funeq}) for the functions $r(q_1+q_2)$ and $s(q_1-q_2)$ from (\ref{sr}).

If $\lambda\ne \pm 1,$ then there exist the following three linearly independent integrals of first degree:
$$
h_1\,=\,e^{-\frac{1}{2} (\lambda+1)\, (q_1+q_2)}\,(p_1+p_2), \qquad h_2\,=\,e^{\frac{1}{2} (\lambda-1)\, (q_1-q_2)}\,(p_1-p_2),\qquad   h=p_2\,-\,\lambda p_1.
$$
Since $h_1\, h_2=-1\, \,({\rm mod}\, H\,=\,0)$, $h_1$ and $h_2$ are functionally dependent, but  the integrals $H$, $h$ and $h_1$ are functionally independent at $H\,=\,0$ for generic $\lambda.$ One can verify that they satisfy the following Poisson algebra on the level $H\,=\,0$:
$$
\{h,\, h_1  \}\,=\,\frac{\lambda^2-1}{2}\,h_1, \qquad \{H ,\, h_1  \}\,=\,
\{H,\, h  \}\,=\,0.
$$

The case $\lambda\,=\,\pm 1$ is degenerate since the integrals $H$, $h,$ and $h_1$ become functionally dependent. However a new additional linear integral  appears just for these degenerate values of $\lambda$. If $\lambda=1,$ then we have
$$h_3\,=\,(p_1\,-\,p_2) \,(q_1\,-\,q_2)\,-\,2\, p_2.$$
 The integrals $H$, $h$, and $h_3$ form a closed Poisson algebra
\begin{equation}\label{dynsym}
\{h ,\, h_3  \}=2 \,h, \qquad \{H ,\, h_3  \}=2 \,H, \qquad \{H ,\, h  \}=0
\end{equation}
off shell of the constraint $H=0$.
A peculiarity of the case $\lambda\,=\,1$   was discussed in details in \cite{FSS} (see also references therein) at the level of phase portraits of the corresponding dynamical systems, which crucially change behavior along the transition line from the range $0<\lambda<1$ to the range $1<\lambda<+\infty$ passing through the critical value $\lambda\,=\, 1$. The restructuring
of the integrals of motion uncovered above and the appearance of the dynamical symmetry (\ref{dynsym}) off shell of the constraint $H=0$ at this value of $\lambda$ shed light on the mathematical origin of that remarkable phenomenon
important in the context of cosmological applications.

\section{Hamiltonians with quadratic integrals}

\subsection{Master equation and its general solution }

Subtracting a polynomial in $p_1, \,p_2$ proportional to $H$, we may write any quadratic additional integral at the level $H=0$ as
\begin{equation}\label{QuadrInt2}
 I_2\,=\,A(q_1,q_2) \, p_1 p_2\,+\,B(q_1,q_2)\, p_2^2+ C(q_1,q_2)\,p_1\,+\,D(q_1,q_2)\,p_2\,+\,
 F(q_1,q_2).
\end{equation}
The condition $\{H,I_2\}=0\, ({\rm mod}\, H=0)$ is equivalent to the following relations
 \begin{equation}\label{AB}
 \frac{\partial A}{\partial q_2}=\frac{\partial B}{\partial q_1}, \qquad \frac{\partial B}{\partial q_2}=\frac{\partial A}{\partial q_1},
 \end{equation}
  \begin{equation}\label{FF}
2 \frac{\partial F}{\partial q_1}-A\,\frac{\partial U}{\partial q_2}=0, \qquad 2 \frac{\partial F}{\partial q_2}+A\,\frac{\partial U}{\partial q_1}+2 B\, \frac{\partial U}{\partial q_2}+2 \frac{\partial A}{\partial q_1}\, U=0,
 \end{equation}
 \begin{equation}\label{con12}
\frac{\partial C}{\partial q_1}=\frac{\partial D}{\partial q_2}, \qquad \frac{\partial C}{\partial q_2}=\frac{\partial D}{\partial q_1},
 \qquad C \frac{\partial U}{\partial q_1}+D \frac{\partial U}{\partial q_2}+2 \frac{\partial C}{\partial q_1}\,U=0.
\end{equation}
 Their simple inspection shows that system (\ref{AB})-(\ref{con12}) is decoupled: equations (\ref{AB}), (\ref{FF}) mean that
 $A\,p_1 p_2+B\, p_2^2+F $ is an integral of second degree while (\ref{con12}) precisely reproduces the set of equations (\ref{con1}), (\ref{con2}) providing the existence of the linear integral $\,C\,p_1+D\,p_2$. Therefore, without loss of generality, one can reduce the ansatz (\ref{QuadrInt2}) putting $C=D=0$ there.

Finding $\frac{\partial F}{\partial q_1}$ and $\frac{\partial F}{\partial q_2}$ from (\ref{FF}) and calculating the compatibility condition $\frac{\partial }{\partial q_2}\frac{\partial F}{\partial q_1}=\frac{\partial }{\partial q_1}\frac{\partial F}{\partial q_2},$ we get
$$
A\,\frac{\partial^2 U}{\partial q_1^2}+2 B\,\frac{\partial^2 U}{\partial q_1 \partial q_2}+A\,\frac{\partial^2 U}{\partial q_2^2}+
3 \frac{\partial A}{\partial q_1} \frac{\partial U}{\partial q_1}+\Big(\frac{\partial A}{\partial q_2}+2 \frac{\partial B}{\partial q_1}  \Big)\,\frac{\partial U}{\partial q_2}+2\,\frac{\partial^2 A}{\partial q_1^2}\,U=0.
$$
For Hamiltonian (\ref{HHour}) the latter condition can be written in the form
\begin{equation}\label{genqv}
 A\, V''(q_2)+\Big(3 \frac{\partial B}{\partial q_1}+ 2B \Big)\,V'(q_2)+ \Big(2 \frac{\partial^2 A}{\partial q_1^2}+3 \frac{\partial A}{\partial q_1}+ A \Big)\,V(q_2)=0.
 \end{equation}
Solving (\ref{AB}), we get (\ref{sr}).
Relation (\ref{genqv}) is a functional equation with respect to $V(q_2),\, r(q_1+q_2)$ and $s(q_1-q_2).$

{\bf Proposition 2.} The potential $V(q_2)$ satisfies the
following linear differential equation
 \begin{equation} \label{EQ}
V''+(2+3 \mu)\,\, \frac{S(q_2)} {C(q_2)}\, V'+ (1+\mu) (1+2 \mu)\, V =0,
 \end{equation}
where  $S(q_2)$ and $C(q_2)$ are given by (\ref{ch}), $\mu,\,a$ and $b$ are constants.

{\bf Proof.} The idea of the proof is the same as in the case of Proposition 1. For fixed $q_2$ relation (\ref{genqv}) is a linear ODE  for $A(q_1,q_2)$ and $B(q_1,q_2)$ with constant coefficients. Differentiating (\ref{genqv}) by $q_2$ and eliminating $q_2$-derivatives of $A(q_1,q_2)$ and $B(q_1,q_2)$ with the help of (\ref{AB}), we get several more linear ODEs. Hence $A(q_1,q_2)$ and $B(q_1,q_2)$ have the form (\ref{AAA}). Substituting these expressions into (\ref{genqv}) and equating coefficients at $q_1^{m_{j}} e^{\lambda_j q_1} $ , we get (\ref{EQ}) with $\mu=\lambda_j, \,a=a_j,\, b=b_j.$ $\square$

It can be easily verified that if we define $A(q_1,q_2)$ and $B(q_1,q_2)$ by (\ref{sr}), (\ref{rs}), then for any solution of (\ref{EQ}) relation
(\ref{genqv}) holds and therefore Hamiltonian $H$ possesses the quadratic integral (\ref{QuadrInt2}).

If a potential satisfies equation (\ref{EQ}), then equations (\ref{FF}) for the function $F(q_1,q_2)$   can be solved  explicitly:
\begin{eqnarray} \label{EQforF}
&& F(q_1,q_2) \,=\,\frac{1}{\mu\,+\,1}\,e^{(\mu\,+\,1)\,q_1}\,C(q_2)\,V^{'} \qquad \mbox{if} \qquad \mu\,\neq \,-1,\\
&& F(q_1,q_2) \,=\,\Big(q_1\,+\, \ln C(q_2)\Big)\,C(q_2)\,V^{'}\,+\,S(q_2)\,V, \qquad \mu\,=\,-1.
\end{eqnarray}
Here $C$ and $S$ are defined by (\ref{ch}).

Let us describe all solutions of (\ref{EQ}) up to the admissible transformations (\ref{tr1}), (\ref{tr2}). In the generic case $a b\ne 0$, $\mu \ne 0,-1, -\frac{1}{2}$   we can normalize $b$ to $\pm a$ by shift (\ref{tr1}).

For the normalization  $b=-a$  we have
\begin{equation} \label{sol1}
 V(q_2)\,=\,c_1 \,\Big(ch \,\frac{\mu q_2}{2} \Big)^{-\frac{2\, (1+\mu)}{\mu}}\,+\,c_2\, \Big(sh \frac{\mu q_2}{2} \Big)^{-\frac{2 \,(1+\mu)}{\mu}}.
 \end{equation}
Notice that the potentials (\ref{deg1}) found in Section 2 are special cases of (\ref{sol1}).

For another possible normalization $b=a$ we find
 \begin{equation} \label{sol2}
 V(q_2)\,=\,c_1 \frac{\cos\Big(\frac{2\, (1+\mu)}{\mu} \arctan e^{\mu q_2}  \Big)}{(ch\, \mu q_2)^{\frac{1+\mu}{\mu}}} \,+\,
 c_2 \frac{\sin\Big(\frac{2\, (1+\mu)}{\mu}
 \arctan e^{\mu q_2}  \Big)}{(ch\, \mu q_2)^{\frac{1+\mu}{\mu}}}.
 \end{equation}
If $\mu=\frac{1}{m},\, (m\in \mathbb{Z})$, this potential can be written as a rational function of hyperbolic functions $ch(\frac{q_2}{m})$ and $sh(\frac{q_2}{m})$.

Consider the special cases  $\mu=0,$  $\mu=-1,$ $\mu=-\frac{1}{2},$ and $a=0.$
In the case $\mu=-\frac{1}{2}$ we get
\begin{equation} \label{sol4}
 V(q_2)\,=\,c_1\,+\,c_2 \Big(a \,e^{-\frac{q_2}{2}}\,-\, b \,e^{\frac{q_2}{2}}\Big).
\end{equation}
The case $\mu=-1$   yields up to involution (\ref{tr2})
\begin{equation} \label{sol5}
V(q_2)\,=\,c_1 \,+\,c_2\, e^{q_2}, \quad V(q_2)\,=\,c_1 \,+\,c_2\, \arctan\Big(e^{q_2}\Big), \quad V(q_2)\,=\,c_1 \,+\,c_2\, \ln\Big(  \frac{e^{q_2}-1}{e^{q_2}+1}  \Big).
\end{equation}
If $\mu=0,$ we get
\begin{equation} \label{sol6}
 V(q_2)\,=\,(c_1\,+\,c_2 \,q_2)\, e^{q_2}, \qquad V(q_2)\,=\,c_1\, e^{\gamma \,q_2}\,+\,c_2 \,e^{\frac{q_2}{\gamma}},
  \end{equation}
 where
 $$
 \gamma^2+2 \frac{a-b}{a+b} \, \gamma+1=0.
 $$
 Notice that for $a b > 0$ the parameter $\gamma$ is a complex number  and therefore we have the following real potential
 \begin{equation} \label{sol7}
 V(q_2)\,=\, e^{\frac{b-a}{a+b}\, q_2}\, \Big(  c_1\,\cos{(\frac{2 \sqrt{a b}}{a+b}\,  q_2)}\,+\,c_2\,\sin{(\frac{2 \sqrt{a b}}{a+b} \, q_2)}\Big).
  \end{equation}
  If $a=b$,  we obtain the potential $V(q_2)=c_1 \cos{q_2}+c_2 \sin{q_2}$, which is related to the $D_2$-Toda \cite{perelomov}.

 The case $a=0$ produces
 \begin{equation} \label{sol8}
 V(q_2)\,=\,c_1 \, e^{(\mu+1)\,q_2}\,+\,c_2\, e^{(2 \mu+1)\,q_2}.
  \end{equation}
The case $b=0$ can be derived from the case $a=0$ by the involution (\ref{tr2}).

\subsubsection{Generalized H\'enon-Heiles and Hietarinta potentials }

In \cite{FSS} several known integrable Hamiltonians with additional integrals of  second degree  were written in the form (\ref{HHour}).  The corresponding potentials for five generalized H\'enon-Heiles Hamiltonians (see \cite{FSS} for details) are given by
 \begin{equation} \label{spor1}
 V(q_2)\,=\,ch(\frac{3 q_2}{5})\,-\,3 \,ch(\frac{q_2}{5}),
  \end{equation}
   \begin{equation} \label{spor2}
  V(q_2)\,=\,ch(\frac{3 q_2}{5})\,+\,7\, ch(\frac{q_2}{5}),
  \end{equation}
    \begin{equation} \label{spor3}
  V(q_2)=3\,-\,ch(\frac{2 q_2}{3}),
  \end{equation}
    \begin{equation} \label{spor4}
  V(q_2)\,=\,63\,-\,60\, ch(\frac{q_2}{3})\,+\,5\, ch(\frac{2 q_2}{3}),
  \end{equation}
    \begin{equation} \label{spor5}
V(q_2)\,=\,63\,+\,60\, ch(\frac{q_2}{3})\,+\,5\, ch(\frac{2 q_2}{3})
  \end{equation}
while for the Hietarinta Hamiltonian the potential has the form
 \begin{equation} \label{sol3}
 V(q_2)\,=\frac{3\, ch (q_2)\,-\,5}{ch^6(\frac{q_2}{2})}\,.
  \end{equation}
According to Proposition 2, these potentials should satisfy (\ref{EQ}) for some $a,\,b,\,\mu$. Indeed, (\ref{spor1}) is given by (\ref{sol2}) with $\mu=-\frac{2}{5},$ $c_1=c_2=\sqrt{2}$. Formula (\ref{sol2}) with $\mu=-\frac{1}{3}, \,c_1=-2, \, c_2=0$ produces  (\ref{spor3}).
Potential (\ref{spor4}) is given  (\ref{sol2})  with $\mu=-\frac{1}{6},\, c_1=0, \, c_2=-8$. Potentials (\ref{spor2}) and (\ref{spor5}) are given by (\ref{sol1}) with $\mu=-\frac{1}{5}, \,c_1=8,\, \,c_2=-8$ and with $\mu=-\frac{1}{6},\, c_1=128,\, c_2=-128,$ respectively.
The Hietarinta potential (\ref{sol3}) is reproduced from (\ref{sol2})
at $\mu=\frac{1}{2}$ and $c_2=0$. For arbitrary $c_1, c_2$ the potential has the form
$$
 V(q_2)\,=\,c_1 \frac{3\, ch (q_2)\,-\,5}{ch^6(\frac{q_2}{2})}\,+\,
 c_2 \frac{sh\, (\frac{3 q_2}{2})\,+\,15\, sh (\frac{q_2} {2})}{ch^6(\frac{q_2}{2})}.
$$

\section{Hamiltonians with cubic integrals}

\subsection{Master equation}

Without loss of generality, any cubic integral at the level $H=0$ can be written in the following form:
\begin{equation}\label{eqthird}
 I_3\,=\,A(q_1,q_2) \, p_1\, p_2^2\,+\,B(q_1,q_2)\, p_2^3\,+\, C(q_1,q_2)\,p_1\,+\,D(q_1,q_2)\,p_2.
\end{equation}
Actually, initially one should also add  second and zero degree terms in momenta to the ansatz (\ref{eqthird}), but it turns out that they form a quadratic integral themselves, so that they can be neglected for the reason that was discussed after equations (\ref{con12}).
Then the condition $\{H,I_3\}=0\, ({\rm mod}\, H=0)$ is equivalent to (\ref{AB}) and to relations
$$
\frac{\partial C}{\partial q_2}-\frac{\partial D}{\partial q_1}+A\, \frac{\partial U}{\partial q_2}=0,
$$
\begin{equation} \label{DC}
C\,\frac{\partial U}{\partial q_1}+D\,\frac{\partial U}{\partial q_2}+2 \, \frac{\partial C}{\partial q_1}\,U=0,
\end{equation}
$$
2\,\frac{\partial C}{\partial q_1}-2\,\frac{\partial D}{\partial q_2}-A \, \frac{\partial U}{\partial q_1}- 3 \, B\,\frac{\partial U}{\partial q_2}-2\,\frac{\partial A}{\partial q_1}\,U  =0.
$$
For Hamiltonian (\ref{HHour}) we get
\begin{equation}\label{tcon3-1}
\frac{\partial \bar C}{\partial q_2}-\frac{\partial \bar D}{\partial q_1}+\bar D+A\,V'=0,
\end{equation}
\begin{equation}\label{tcon3-2}
3\,\bar C\,V+\bar D\,V'+2 \, \frac{\partial \bar C}{\partial q_1}\,V=0,
\end{equation}
\begin{equation}\label{tcon3-3}
2\,\frac{\partial \bar C}{\partial q_1}-2\,\frac{\partial \bar D}{\partial q_2}-A \, V'- 3 \, B\,V'-2\,\frac{\partial A}{\partial q_1}\,V +2 \bar C =0,
\end{equation}
where
$$
\bar C(q_1,q_2)\equiv e^{- q_{1}}\, C(q_1,q_2), \qquad \bar D(q_1,q_2)\equiv e^{- q_{1}}\, D(q_1,q_2).
$$

{\bf Proposition 3.} If Hamiltonian (\ref{HHour}) possesses integral (\ref{eqthird}) on the level $H=0$, then the potential $V(q_2)$ satisfies the
following nonlinear differential equation of third order:
\begin{equation} \label{mastereq}
a_1 \,V V'^2 V'''\,+\,a_2\, V^2 V' V'''\,+\,a_3 \,V V' V''^2\,+\,a_4\, V^2 V''^2\,+\,a_5\, V'^3 V''\,+\,a_6\, V V'^2 V''\,+\,
\end{equation}
$$
a_7 \,V^2 V' V''\,+\,a_8\, V^3 V''\,+\,a_9\, V'^4\,+\,a_{10} \,V V'^3\,+\,a_{11} \,V^2 V'^2\,+\,a_{12}\, V^3 V'\,+\,a_{13} \,V^4\,=\,0,
$$
where
$$
a_1\,=\,-6 (3+2 \mu)\,  S(q_2), \quad
a_2\,=\,-(14+12 \mu)(3+2 \mu)\, C(q_2),\quad
a_3\,=\,18 (3+2 \mu)\,S(q_2),\quad
$$
$$
a_4\,=\,16 (1+\mu) (3+2 \mu)\,C(q_2), \qquad
a_5\,=\, -6 (7+5 \mu)\, S(q_2),\qquad
a_6\,=\,2 (3+2 \mu) (16+15 \mu)\, C(q_2),
$$$$
a_7=2 (3+2 \mu) (4+3 \mu) (9+10 \mu)\, S(q_2),\qquad
a_8=2 (1+\mu) (3+2 \mu)^2 (9+10 \mu)\, C(q_2),\qquad
$$
$$
a_9=-30 (1+\mu) (4+3 \mu)\, C(q_2), \qquad
a_{10}=-2 (90+241 \mu+210 \mu^2+60 \mu^3)\, S(q_2),
$$
$$
a_{11}\,=\,-2 (1+\mu) (3+2 \mu)\, C(q_2),\qquad
a_{12}\,=\,2 (1+\mu) (3+2 \mu)^2 (9+22 \mu+12 \mu^2)\, S(q_2),
$$
$$
a_{13}\,=\,2(1+\mu)^2 (1+2 \mu) (3+2 \mu)^3\,  C(q_2).
$$
Here the functions $C(q_2)$ and $S(q_2)$ are given by (\ref{ch}). We will call (\ref{mastereq}) the {\it master equation}.


{\bf Proof.} The proof is similar to the ones of Propositions 1 and 2. Differentiating (\ref{tcon3-2}) several times  by $q_2$ and eliminating $q_2$-derivatives of $\bar C,\, \bar D,\, A$  and $B$ with the help of (\ref{tcon3-1}), (\ref{tcon3-3}) and (\ref{AB}), we arrive at a system of linear equations with respect to the independent variable $q_1$. This implies that $A(q_1,q_2)$ and $B(q_1,q_2)$ are quasi-polynomials of the form (\ref{AAA}) and
\begin{equation}\label{CD}
\bar C(q_1,q_2)\,=\,\sum \bar C_j(q_1,q_2) e^{\lambda_j q_1} , \qquad \bar D(q_1,q_2)\,=\,\sum \bar D_j(q_1,q_2) e^{\lambda_j q_1} , \qquad
\end{equation}
where $\bar C_j(q_1,q_2)\,=\,\bar c_j(q_2) \,q_1^{m_{j}}+\cdots,\quad \bar D(q_1,q_2)\,=\,\bar d_j(q_2)\, q_1^{m_{j}}+\cdots$ are polynomials in $q_1$.  Substituting these
expressions for  $\bar C(q_1,q_2), \,\bar D(q_1,q_2),\, A(q_1,q_2)$  and $B(q_1,q_2)$ into (\ref{tcon3-1})--(\ref{tcon3-3}) and equating coefficients at $q_1^{m_{j}} e^{\lambda_j q_1} $ , we get the following {\it master system}:
 $$
 P'=(1+\mu) Q- \frac{1 + 2 \mu}{2}C(q_2)\, V-\frac{3}{2} S(q_2)\, V',
 $$
 \begin{equation}\label{master}
 Q'=(1+\mu) P- C(q_2)\, V',
 \end{equation}
 $$
P\, V'=- (3+2 \mu) Q  \, V,
 $$
 where $\mu=\lambda_j,\, a=a_j,\, b=b_j,$ $P(q_2)=d_j(q_2),\, Q(q_2)=c_j(q_2)$.
 Eliminating the functions $P(q_2)$ and $Q(q_2)$ from (\ref{master}), we arrive at (\ref{mastereq}). $\square$

 {\bf Remark 1.}  Master system (\ref{master}) and master equation (\ref{mastereq}) admit the following involution: $$q_2\to -q_2,\quad a\to b, \quad b\to a,\quad V\to -V,\quad P\to P,\quad Q\to -Q.$$

 To show that equation (\ref{mastereq}) is a sufficient condition for integrability, let us put
 $$
P(q_2)\,=\,\frac{(2 \mu+3) \,V V'\, \Big((6 \mu + 7)\, C(q_2)\,  V\,+\, 3\, S(q_2)\, V'\Big)}{(2 \mu+3)\, V V''\,-\,(3 \mu +4)\, V'^2\,+\,(\mu+1) (2 \mu+3)^2 \, V^2},
$$
$$
Q(q_2)\,=\,-\frac{V'^2 \,\Big((6 \mu + 7) \,C(q_2)\,  V\,+\, 3\, S(q_2)\, V'\Big)}{(2 \mu+3)\, V V''\,-\,(3 \mu +4)\, V'^2\,+\,(\mu+1) (2 \mu+3)^2\, V^2},
$$

 $$
 A(q_1,q_2)\,=\,e^{\mu q_1}\, C(q_2), \qquad B(q_1,q_2)\,=\,e^{\mu q_1}\,S(q_2),
 $$
 $$
 D(q_1,q_2)\,=\,e^{(\mu+1) q_1}\, P(q_2), \qquad C(q_1,q_2)\,=\,e^{(\mu+1) q_1}\, Q(q_2).
 $$
 Then one can verify that relations (\ref{AB}), (\ref{DC}) are fulfilled in virtue of (\ref{mastereq}).

{\bf Remark 2.}  The denominator in expressions for $P(q_2)$ and $Q(q_2)$ vanishes if and only if the potential $V(q_2)$ satisfies (\ref{eqlin}), so that the corresponding Hamiltonian admits a linear integral of the form (\ref{lin}). Indeed, the function $y(q_2)=\frac{C(q_2)}{S(q_2)},$
where $\mu\to \nu$,  satisfies the ODE $y'+\nu y^2-\nu=0.$ It follows from (\ref{eqlin}) that $y=-\frac{V'}{(2\nu+1)\,V}.$ Substituting it into the ODE, we get the denominator (up to a nonzero factor and to the replacement $\nu\to \mu+1$).

 {\bf Remark 3.}  In the case $\mu=-\frac{3}{2}$ equation (\ref{mastereq}) has a particular solution $V=const$. For nonconstant solutions equation (\ref{mastereq}) reduces to a linear equation of second order, which
coincides   with (\ref{EQ}), where  $\mu=-\frac{3}{2},$  up to $a\to b, b\to -a$. For this reason we assume below that  $\mu\ne -\frac{3}{2}.$

\subsubsection{Especial solutions in the case  $a b\neq 0$}


In the future we plan to investigate master equation (\ref{mastereq}) with   $a b\ne 0$.
By now  we have found several its explicit solutions   for special values of the parameter $\mu$.

{\bf Example 4.1}.  Let $\mu=0$. In this case the ratio $a b^{-1}$ is an essential parameter.  If $a b^{-1}=1,$ then the general solution written in a parametric form is given by
 $$
 V(x)\,=\,\frac{(F')^{
 \frac{3}{2}}}{F}, \qquad \frac{d q_2}{d x}\,=\,2 \sqrt{3}\, \frac{F^{\frac{1}{2}}}{F'},
 $$
 where
 $$F(x)\,=\,3\, \varepsilon^2\,+\,\delta\, x\,+\,6\, \varepsilon\, x^2\,-\,x^4$$
 and $\delta$ is an arbitrary constant. If $\varepsilon\ne 0,$ then it can be reduced to $1$ by a rescaling of the parameter $x$. Two more arbitrary constants in the solution can be reconstructed by transformation (\ref{tr1}). This transformation is admissible  since for $\mu=0$ equation (\ref{mastereq}) does not depend on $q_2$ explicitly. The corresponding Lagrangian coincides with the integrable case (22) from \cite{yehia}, where $\alpha=-2, \quad \mu=-4, \quad \nu=x, \quad p_0=h_1=A_2=0.$

{\bf Example 4.2}. If $\mu=0$ and $a b^{-1}=-1,$ the general solution can be written as
 $$
 V(x)\, =\, x^{\frac{3}{2}}, \qquad \frac{d q_2}{d x}= \frac{\sqrt{3}}{2}\ \frac{x^2-\varepsilon}{x\, \sqrt{F(x)}}.
 $$

If $\varepsilon = 0$, then  $q_2(x)$ in Examples 4.1 and 4.2 can be expressed in terms of elementary functions. This is also true for $\varepsilon=1,\, \delta=\pm 8.$ In particular, Example 4.2 with $\varepsilon = 0$ leads to the potential $V(q_2)\,=\,\sin(\sqrt{3}\, q_2)$
related to the $A_2$-Toda \cite{FSS}.

 {\bf Example 4.3}.  If $\mu=-2$ and $a=\pm b=1,$ then there exists the following particular solution in hypergeometric functions:
$$
V(q_2)\,=\,c_1\, e^{\frac{q_2}{3}}\,
   F\Big(\frac16, \frac56; \frac13; \mp e^{4 q_2}\Big)\, +\, c_2\, e^{3 q_2}\,
   F\Big(\frac56, \frac32; \frac53; \mp e^{4 q_2}\Big).
$$

For the sake of completeness, we also present several earlier known examples of integrable Hamiltonians with cubic integrals (see \cite{FSS} and references therein) written in the form (\ref{HHour}) in \cite{FSS}.

{\bf Example 4.4}. The Fokas--Lagestrom--Inozemtsev potential. Under the normalization $b=-a$ the potential
$$
V(q_2)\,=\,ch^{-\frac{2}{3}}(3 q_2)
$$
satisfies (\ref{mastereq}) with $\mu=-3.$

{\bf Example 4.5}. The Holt--Drach potential
$$
{\displaystyle V(q_2)\,=\,\frac{ch(\frac{3 q_2}{5})\,-\,7}{ch^{\frac{2}{3}}(\frac{3 q_2}{10})} }
$$
satisfies  (\ref{mastereq}) with  $b=-a$ and $\mu=-\frac{9}{10}$.

{\bf Example 4.6}. The Drach potential
$$
{\displaystyle V(q_2)\,=\,\frac{sh(\frac{3 q_2}{7})}{ch^{\frac{2}{3}}(\frac{3 q_2}{7})} }
$$
satisfies  (\ref{mastereq}) with  $b=-a$ and $\mu=-\frac{9}{7}$.

{\bf Example 4.7}.
The Galajinsky-Lechtenfeld potential \cite{galech}
\begin{eqnarray}\label{SashaGL}
 V(q_2)\,=\,\left(3\, sh^2(\frac{q_2}{4})\,  ch(\frac{q_2}{4})\,+\, ch^3(\frac{q_2}{4})\right)^{-\frac{2}{3}}\,
\end{eqnarray}
satisfies (\ref{mastereq}), where  $\mu=-\frac{7}{4}$ and the constants $a$ and $b$ are arbitrary. Notice that this potential coincides with  (\ref{deg1}),  where $\mu=-\frac{3}{4}$.  Therefore the corresponding Hamiltonian possesses a first degree integral as well.

\subsection{Case $b=0$. Linearization of master equation and its general solution}

Suppose that $b=0, \,a=1$ (see Remark 1)  and $\mu \neq -\frac{3}{2}$ in (\ref{ch}), (\ref{mastereq}), (\ref{master}).
One can verify that master system (\ref{master}) possesses the following integral of motion in this case :
\begin{eqnarray}\label{Sasha01}
k\,\equiv\,\frac{1}{4}\, e^{(2\, \mu\,+1) \,q_2} V(q_2)\,+\,\frac{1}{2}\, e^{(\mu\,+\,1) \,q_2}
   \Big(P(q_2)\,-\,Q(q_2)\Big)\,, \qquad \frac{d}{d q_2}k\,=\,0\,.
\end{eqnarray}
Furthermore, excluding $P$ and $Q$ from (\ref{master}) with the use of (\ref{Sasha01}), we get a second order ODE for  the function
\begin{eqnarray}\label{Sasha02a}
z(q_2)\,\equiv\, e^{(2 \mu \,+\,1)\,q_2} V(q_2)
\end{eqnarray}
that does not depend on $q_2$  explicitly. Hence its order can be reduced to one by the substitution
\begin{eqnarray}\label{Sasha02}
\frac{d}{d q_2}\,z(q_2)\,=\,g(z(q_2))\,.
\end{eqnarray}
As a result we arrive at the following Abel equation of the second kind:
\begin{eqnarray}\label{Sasha1}
&& g(z)\,\frac{d \,g(z)}{d z}\,+\,f_0(z)\,+\,f_1(z)\,g(z)\,+\,
f_2(z)\,g(z)^2+\,f_3(z)\,g(z)^3\,=\,0\,,\qquad \mbox{where}\\
&& f_0(z)\,=\,\frac{8\, z\, \Big(\mu ^2 \,z\,-\,4\, k\, (\mu +1)^2\Big)}{(2 \mu
   +3) (z \,-\,4 k)}\,,\nonumber\\
&& f_1(z)\,=\,\frac{2 \,\Big( (2 \mu -7)\,\mu\,
   z\,-\,4\, k\, (\mu +1)\, (2 \mu +1)\Big)}{(2\, \mu +3) \,(z \,-\,4 k)} \,, \nonumber\\
&& f_2(z)\,=\, \frac{4 \,k\, (3 \,\mu + 4)\,-\,3 \,(3\, \mu -1) \,z}{(2 \mu +3)\, z\, (z \,-\,4\,k)}\,, \nonumber\\
&& f_3(z)\,=\, \frac{3}{(2 \mu +3)\, z \,(z \,-\,4 k)}\,. \nonumber
\end{eqnarray}
By the transformation $\{z,~g\}\,\rightarrow\, \{\tau, ~x\}$ of the form
\begin{eqnarray}\label{Sasha2}
z\,=\,\tau^2, \qquad g\,=\,
2\,\tau^2\,\frac{\omega_1 \,\tau^2\,+\, x \,\tau\,+\, k\,\omega_2}{(2\,-\,\omega_1)\, \tau^2\,-\, x \,\tau\,-\,k\,(8\, + \,\omega_2)}\,
\end{eqnarray}
the Abel equation can be simplified:
\begin{equation}\label{ric0}
\,(\omega_1\, \tau^2\,+\, x \,\tau\,+\,k\,\omega_2)\,\frac{d x}{d \tau}=f(x),
\end{equation}
where
$$
 \omega_1\,\equiv\,\frac{4\, \mu}{2 \mu\,+\,3}\,, \qquad \omega_2\,\equiv \,-\frac{16 \,(\mu\,+\,1)}{2 \mu\,+\,3}\,, \qquad
f(x)\,\equiv\,-(2 \mu+3) \,x^2\,+\,\frac{64 \, k}{2\,\mu\,+\,3}\,.
$$

A simple inspection of equation (\ref{ric0}) shows that it has a particular solution
\begin{equation}\label{ric0sol}
x\,=\,\pm\,\frac{8 \, \sqrt{k}}{2\,\mu\,+\,3},
\end{equation}
which being substituting into (\ref{Sasha2})
generates the following parametric representation of the potential (\ref{Sasha02a}), (\ref{Sasha02}):
\begin{eqnarray}\label{Sasha02ParamRic}
e^{q_2(\tau)}&=&\tau ^{\frac{1}{2\, (\mu\,+\,1)}}
\left(2 \,\sqrt{k} \,(\mu\,+\,1)\,\pm \,\mu\,
\tau \right)^{\frac{2 \,\mu\,+\,3}{2\, \mu  \,(\mu\,+\,1)}}\,, \nonumber\\
V(\tau)&=&\tau ^{\frac{2\,\mu\,+\,3}{2\,(\mu\,
+\,1)}} \left(2\,\sqrt{k}\,(\mu\,
+\,1)\,\pm\,\mu\, \tau\right)^{-\frac{(2\,\mu\,+\,1)\,
(2\,\mu\,+\,3)}{2\,\mu\,(\mu\,+\,1)}}
\end{eqnarray}
with the parameterizing parameter $\tau$.
Excluding $\tau$, one can derive a quasi-algebraic relationship between $V$ and $e^{q_2}$ for this case:
\begin{eqnarray}\label{eqVAnyMu}
c_3\,+\,c_2\, e^{\frac{2\,\mu\,+\,1}{2}\,q_2}\, V^{\frac{1}{2}}\,+\,c_1\, e^{\frac{\mu}{2}\,q_2}\,V^{-\frac{\mu}{2\,(2\,\mu\,+\,3)}} \,=\,0\,.
\end{eqnarray}
Hereafter we use a freedom (\ref{tr1}) to rescale the potential $V(q_2)$ and to shift the coordinate $q_2$ as well as to multiply an equation by a constant to reconstruct arbitrary constant parameters $c_1,\, c_2,\,c_3$ in (\ref{eqVAnyMu}).

The limit $\mu\to 0$ in (\ref{Sasha02ParamRic}), together with a rescaling of $\tau$, yields the following particular solution:
\begin{equation}\label{eq1}
V=\tau^{\frac{3}{2}} \, e^{-\tau},  \qquad e^{q_{2}}= \tau^{\frac{1}{2}}\, e^\tau,
\end{equation}
which will be discussed in what follow (see the discussion after Example 4.11).

Written in the form
\begin{equation}\label{ric}
f(x)\,\frac{d \tau}{d x}\,=\,\omega_1\, \tau^2\,+\, x \tau\,+\, k\,\omega_2
\end{equation}
 equation (\ref{ric0}) becomes a Riccati equation with respect to the function $\tau(x)$, which is related to the associated Legendre equation (see eq. (1.3.3)-(2.11) in \cite{abel}).

Given a solution $\tau(x),$ we obtain the parametric representation of the potential (\ref{Sasha02a}), (\ref{Sasha02})
\begin{eqnarray}\label{Sasha02Param}
V(x)\,=\, e^{-(2 \,\mu \,+\,1)\,q_2(x)} \,\tau(x)^2 \,,
\quad q_2(x)\,=\,\int \frac{dx}{f(x)}\, \Big(\frac{6}{2\,\mu\,+\,3}\,\tau(x)-x-\frac{8\,k}{(2\, \mu\,+\,3)\,\tau(x)}\Big),
\end{eqnarray}
where $x$ plays the role of a parameter.

In the generic case\footnote{We recall that $\mu \,\neq\,-\frac{3}{2}$ in the case under consideration.} $ k\,\neq \,0\,,k\,\neq \infty,\, \mu \,\neq\, 0$ the general solution of (\ref{ric}) can be expressed in terms of the associated Legendre functions:
\begin{equation}\label{gensol}
\tau(x)= \frac{2\sqrt{k} (\mu +3)}{\mu} \,
\frac{c\, P_{\alpha +1}^{\beta}(\rho )+
Q_{\alpha +1}^{\beta }(\rho )}{c\,P_{\alpha }^{\beta }(\rho )+
Q_{\alpha }^{\beta}(\rho )}-\frac{(\mu +2) (2 \mu+3)}{2\mu}\,\,x,
\end{equation}
where
$$
\alpha\,\equiv\, \frac{1}{2\,(3\,+\,2\,\mu)}\,,
\qquad \beta\,\equiv\, \frac{1\,+\,2 \,\mu}{2\,(3\,+\,2\, \mu)}\,,\qquad \rho\,\equiv\, \frac{3\,+\,2 \,\mu}{8\, \sqrt{k}}\, x
$$
and $c$ is an arbitrary constant.

Consider the special cases. If $k=0$ and $\mu\ne -2$ the general solution is given by
$$
\tau(x)= \frac{(\mu \,+\,2) (2\, \mu\,+\,3)\, x}{c\, x^{\frac{2\,(\mu\,+\,2)}{2\,\mu\,+\,3}}\,
-\,2\,\mu}.
$$
Substituting $\tau(x)$ into equations (\ref{Sasha02Param}), we find that modulo of an inessential constant factors
\begin{eqnarray}\label{SashaPot1}
e^{q_2(x)}&=&x^{\frac{2}{\mu} } \left(c \,x^{\frac{2 (\mu \,+\,2)}{2\, \mu\, +\,3}}\,-\,2 \,\mu \right)^{-\frac{3}{2\, \mu }}\,,\nonumber\\
V(x)&=& x^{-\,\frac{2\, (\mu\, +\,1)}{\mu}} \left(c\, x^{\frac{2 \,(\mu \,+\,2)}{2\, \mu \,+\,3}}\,-\,2\, \mu
   \right)^{\frac{2\, \mu\, +\,3}{2\, \mu }}.
\end{eqnarray}
Excluding parameterizing parameter $x,$ we obtain a quasi-algebraic relationship between $V$ and $e^{q_2}$:
\begin{eqnarray}\label{eqVversusQ2k0}
c_1\,+\,c_2\, e^{\mu\,q_2} V^{\frac{\mu}{2\,
\mu\,+\,3}}\,+\,c_3\, V^2 e^{2\,(\mu\,+\,1)\,q_2}\,=\,0\,.
\end{eqnarray}

In the case  $k\,= \,0,\, \mu \,=\, -2$ we have
$$
\tau(x)=\frac{x}{c\,-\,8 \ln (x)}.
$$

For $\mu=0$ the general solution can be expressed in terms of the hypergeometric function:
$$
\tau(x)= \frac{c\,-\,x \,F\left(\frac{1}{2},\frac{5}{6};\frac{3}{2};
   \frac{9\, x^2}{64\, k}\right)}{4\,(1\,-\,\frac{9\, x^2}{64\, k})^{\frac{1}{6}}}.
$$

In the limit $k=\infty$ transformation (\ref{Sasha2}) is not invertible. After the rescaling $x\,\rightarrow\,k\,x$ in (\ref{Sasha2}) we obtain
the correct transformation $\{z,~g\}\,\rightarrow\, \{\tau, ~x\}$ for  $k=\infty$:
\begin{eqnarray}\label{Sasha2corr}
z\,=\,\tau^2, \qquad g\,=\,
-2\,\tau^2\,\frac{x \,\tau\,+\, \omega_2}{\, x \,\tau\,+\,(8\, + \,\omega_2)}\,.
\end{eqnarray}
The corresponding equation
\begin{equation}\label{ric0corr}
-(2 \mu+3) \,x^2\,=\,(x \tau\,+\,B)\,\frac{d x}{d \tau}
\end{equation}
valid at $k\,=\,\infty$. It is a linear equation with respect to $\tau(x)$ whose solution can be easily constructed.
It follows from Remark 2 that in this case there exists an integral of first degree (see Section 2).

\subsubsection{Explicit solutions of master equation}

\qquad In this Subsection we follow the ideas of \cite{kovach} although we cannot straightforwardly apply the algorithm from \cite{kovach}  due to the dependence of equation (\ref{ric}) on the parameter $\mu.$

\qquad
{\bf Theorem 1.} Equation (\ref{ric}) has a rational solution $\tau(x)$ iff $\mu$ belongs to one of the following six families:
\begin{equation}
\begin{array}{ccc}\label{mu}
{\bf 1:}\quad  \mu\,=\,\frac{1\,-\,6 \,n}{4\, n} \quad (n\ne 0), \qquad {\bf 2:}\quad \mu\,=\,-\frac{7\,+\,6 \,n}{4\, (1\,+\,n)}, \qquad {\bf 3:}\quad \mu\,=\,-\frac{3\, n}{1\,+\,2\, n}, \\[5mm]
\displaystyle {\bf 4:}\quad \mu\,=\,-\frac{1\,-\,3\, n}{1\,-\,2\, n},\qquad ~~~~~~~~~~{\bf 5:}\quad \mu\,=\,-\frac{3\,(1\,+\,n)}{1\,+\,2n}, \qquad {\bf 6:}\quad \mu\,=\,-\frac{4\,+\,3 \,n}{3\,+\,2\, n},
\end{array}
\end{equation}
where $n$ is an arbitrary nonnegative integer.

{\bf Proof.} The transformation
\begin{equation}\label{tauy}
\tau\,=\,\frac{(3\,+\,2\, \mu)^2}{4\, \mu} (x^2\,-\,d^2) \,y\, -\, \frac{(3\,+\,2\, \mu)\, (7\,+\,4\, \mu)}{8 \,\mu} \,x
\end{equation}
reduces the Riccati equation (\ref{ric}) to the canonical form
\begin{equation}\label{canric}
y'\,+\,y^2\,+\,r(x)\,=\,0,
\end{equation}
where
\begin{equation}\label{rr}
r(x)\,=\,\frac{m_2}{(x\,+\,d)^2}\,+\,\frac{m_2}{(x\,-\,d)^2}\,+\,\frac{m_1}{x\,+\,d}\,
-\,\frac{m_1}{x\,-\,d}.
\end{equation}
The parameters in (\ref{ric}) and (\ref{canric}) are related by the following formulas:
\begin{equation}\label{mm}
d\,=\,\frac{8\,\sqrt{k}}{3\,+\,2\, \mu}, \qquad m_2\,=\,\frac{(5\,+\,2\, \mu)\,(7\,+\,6\, \mu)}{16\, (3\,+\,2 \,\mu)^2}, \qquad m_1\,=\,\frac{49\,+\,52\, \mu\,+\,12\, \mu^2}{128\, \sqrt{k} \,(3\,+\,2\, \mu)}.
\end{equation}
Substituting the partial fraction decomposition of an arbitrary rational function into (\ref{canric}), we arrive at the following statement.

{\bf Lemma}. Any rational solution of (\ref{canric}) has the following structure:
\begin{equation}\label{ratsol}
y=\frac{p}{x-d}+\frac{q}{x+d}+\sum_{i=1}^n \frac{1}{x-\gamma_i}. \qquad \square
\end{equation}

Substituting (\ref{ratsol}) into the left hand side of (\ref{canric}), we get a rational function $R(x)$ with the poles at $x=\pm d$ and $x=\gamma_i$.
It follows from (\ref{rr}) and (\ref{ratsol}) that the function $R(x)$ tends to zero as $x\to \infty.$ Therefore $R$ is identically equal to zero iff the principle parts of the Laurent series at these points vanish. This is equivalent to the following relations:
\begin{equation}\label{cond1}
m_2\,=\,p\,-\,p^2\,=\,q\,-\,q^2, \quad m_1\,-\,\frac{p\, q}{d}\,-\, 2 p\, \sum_{i=1}^n \frac{1}{d-\gamma_i}\,=\,0, \quad m_1\,-\,\frac{p\, q}{d}\,-\, 2 q \sum_{i=1}^n \frac{1}{d+\gamma_i}=0
\end{equation}
and
\begin{equation}\label{cond2}
\frac{p}{\gamma_i-d}+\frac{q}{\gamma_i+d}+\sum_{j\ne i}\frac{1}{\gamma_i-\gamma_j}\,=\,0, \qquad i=1,...,n.
\end{equation}
Summing all relations (\ref{cond2}) with different $i,$ we obtain
$$
p \, \sum_{i=1}^n \frac{1}{d-\gamma_i}\,=\,q\, \sum_{i=1}^n \frac{1}{d+\gamma_i}
$$
and consequently the second and the third relations in (\ref{cond1}) coincide.  Therefore relations (\ref{cond1}) and (\ref{cond2}) form a system of $n+3$ algebraic equations with respect to $n+3$ unknowns $\mu,\, p,\, q$ and $\gamma_i.$ For small $n$ this system can be solved explicitly. To derive a closed system of equations for $p,\,q$ and  $\mu,$ we consider
the expansion of $R$ as $x\to \infty$. One can observe that $R\,=\,\sum_{k=2}^{\infty} c_k\, x^{-k}$, where the only coefficient $c_2$ does not depend on $\gamma_i.$  The condition $c_2=0$, together with the first two relations (\ref{cond1}), gives rise to the system of equations
\begin{equation}\label{syssol1}
m_2\,=\,p\,-\,p^2\,=\,q\,-\,q^2, \qquad 2\, p\,q\,+\,2\, n \,(p\,+\,q)\,+\,n^2\,-\,n\,-\,2\, m_1\, d\,=\,0.
\end{equation}
Using (\ref{mm}), we finally derive a system of three equations for the three parameters $p,\,q$ and $\mu$. We see that the system is symmetric with respect to the variables $p$ and $q$. All solutions of these system up to the involution $p \leftrightarrow q$ are presented in the Table 1.

{}~

{}~

\bigskip

\centerline{{\bf Table 1}}

\medskip

\begin{tabular}{|*{7}{p{50pt}|}}\hline \rule{0pt}{15pt} Family & ~~~~1& ~~~~2 &~~~~3&~~~~4&~~~~5&~~~~6\\ \hline

$~~~~\mu$& $~~~~\frac{1-6 n}{4 n}$ & $-\frac{7+6 n}{4 (1+n)}$ & $-\frac{3 n}{1+2 n}   $ & $-\frac{1-3 n}{1-2 n}$ & $-\frac{3(1+n)}{1+2n}$ & $-\frac{4+3 n}{3+2 n}$ \\ \hline

$~~~~p$ &  $~~~\frac{1}{4}+n$ &  $~~\frac{7}{4}+n$ & $\frac{7}{12}-\frac{n}{3}$ & $~~\frac{3}{4}-n$ & $~~\frac{1}{12}-\frac{n}{3}$ & $-\frac{3}{4}-n$
\\ \hline

$~~~~q$ &  $~~~1-p$ &  $~~1-p$ & $~~~~~p$ & $~~~~~p$ & $~~~~~~p$ & $~~~~~p$
\\ \hline
\end{tabular}

\bigskip
\noindent In Table 1 $n$ is an arbitrary nonnegative integer, which is equal to the number of poles $\gamma_i$ in (\ref{ratsol}). If we have no such poles, then $n=0.$ For the first family $n \ne 0.$

\medskip

Given a solution of (\ref{syssol1}), we are to solve the system of nonlinear algebraic equations (\ref{cond1}) and (\ref{cond2}) for $\gamma_i$. It turns out that it  is equivalent to a linear system of equations for the coefficients $q_i$ of the polynomial $Q(x)=(x-\gamma_1)\cdots(x-\gamma_n)\,\equiv\,
x^n\,+\,\sum_{i=0}^{n-1} q_i\,x^i$. Furthermore, the latter system has the following triangular structure:
\begin{equation}\label{lins}
(i-n) \,(i+n-1+2\, p+ 2\, q)\, q_i\,=\,Z_i, \qquad i=n-1,\dots,0,
\end{equation}
where $Z_i$ is a linear function in $q_{i+1},\,q_{i+2},\,q_{i+3}$ and $q_{i+4}$. In particular, $q_{n-1}=(p-q) d$. It follows from Table 1 that the coefficient at $q_i$ in (\ref{lins}) never equals zero.

One can derive relations (\ref{lins}) in the following way.
It follows from (\ref{ratsol}) that the corresponding Schr\"odinger equation
$$\Psi''\,+\,r\, \Psi\,=\,0, \qquad  y\,\equiv\,\frac{\Psi'}{\Psi}
$$
possesses the solution
$$
\Psi\,=\,(x-d)^p \,(x+d)^q\, Q(x).
$$
Substituting such $\Psi$ into the Schr\"odinger equation and multiplying the result by the expression $(x-d)^{p-2} \,(x+d)^{q-2}$, we give rise to
a linear relation between $Q,\, Q'$ and $Q''$. Equating the coefficients at different powers of $x$, we obtain (\ref{lins}).
 $\square$

For the families 1 and 2 from Table 1 we have $q=1-p$ and there exist two rational solutions related by the involution $x\to -x,\, y\to -y, \, p \leftrightarrow q.$
For the  families $3,\, 4,\, 5$ and $6$ we have $p=q$ and there exists only one rational solution invariant with respect to the involution.

The simplest solutions correspond to $n=0$, $Q=1$. It follows from equation (\ref{tauy}) that if $n=0,$ we get solutions of equation (\ref{ric}) linear in $x$.

Without loss of generality, we fix  $k=\frac{1}{64}$ in (\ref{mm}) by a rescaling of $x$. Under such a normalization
\begin{equation}\label{dd}
d=\frac{1}{3+2 \mu}.
\end{equation}

\paragraph{Examples}\mbox{} \\

{\bf Example 4.8}. Consider the family 1. In follows from (\ref{dd}) that $d= 2 n.$ If $n=2$, then $Q\,=\,x^2\,+\,14\, x\,+\,60$, for $n=3$ we have $Q\,=\,x^3\,+\,33\, x^2\,+\,414 \,x\,+\,2079.$

{\bf Example 4.9}. Let $n=1$. Then $Q=x+3$ for the family 1, $Q=x+18$ for the family 2, and $Q=x$ for the families $3,\, 4,\, 5$ and $6$.

{\bf Example 4.10}. Consider the family 1 with $n=1$ in more details. We have $\mu\,=\,-\frac{5}{4}$. Taking $q\,=\,\frac{5}{4}$, we get $$y\,=\,\frac{5}{4} \frac{1}{x\,+\,2}\,-\,\frac{1}{4} \,\frac{1}{x\,-\,2}\,+\,\frac{1}{x\,-\,3}.$$
Formula (\ref{tauy}) gives us
$$
\tau\,=\,-\frac{1}{4\, (x\,-\,3)}.
$$
It follows from (\ref{Sasha02Param}) that
$$
q_2\,=\,\frac{6}{5}\,\ln\Big(\frac{3\,-\,x}{2\,+\,x}  \Big), \qquad V=\frac{1}{16} \,(2\,+\,x)^{-\frac{9}{5}} \, (3\,-\,x)^{-\frac{1}{5}}.
$$
Eliminating $x$, we obtain
$$
V\,=\,\frac{1}{400} \,\Big(e^{-\frac{q_{2}}{12}}\,+\, e^{\frac{3 \,q_{2}}{4}} \Big)^2.
$$
Reconstructing two arbitrary constants by transformation  (\ref{tr1}), we arrive at Hamiltonian
(\ref{new}).

Some of the potentials corresponding to rational solutions with small $n$ have already appeared in Sections 2 and 3. Here we present explicit integrable potentials that are new. All of them have the following parametric form:
\begin{equation} \label{param}
V=c_1\, \tau^{k_{1}} \, (\tau-1)^{k_{2}}, \qquad e^{q_2} = c_2\, \tau^{s_{1}} (\tau-1)^{s_{2}},
\end{equation}
where  $k_i, s_i$ are constants, and $c_i$ are arbitrary parameters. For example, solution  (\ref{SashaPot1}) can be written in this form  after a proper reparameterization.  The formulas for $q_2$  and $V$ from Example 4.10 can also be written in the form (\ref{param}) after the reparameterization $x\,=\,-5\, \tau\,+\,3.$

It can be easily verified that any solution (\ref{param}) satisfies a quasi-algebraic equation of the form  (cf. with   (\ref{eqVAnyMu}), (\ref{eqVversusQ2k0}))
\begin{equation} \label{paramNew1}
\tilde c_1 \, e^{l_1 q_2}\,V^{r_1}\,+\,\tilde c_2\, e^{l_2 q_2}\,V^{r_2}
\,=\,\tilde c_3,
\end{equation}
where
$$
r_1=\frac{s_1}{s_1 k_2-s_2 k_1}, \qquad l_1=-\frac{k_1}{s_1 k_2-s_2 k_1}, \qquad r_2=-\frac{s_2}{s_1 k_2-s_2 k_1}, \qquad  l_2=\frac{k_2}{s_1 k_2-s_2 k_1}.
$$
In Table 2 we present several examples of this kind. For each example we write down the number of the family in (\ref{mu}), the integer $n$, the values of $\mu$,  $k_i,\, s_i$ as well as $r_i,\, l_i$ in (\ref{param}) and (\ref{paramNew1}), respectively. The constants  $k_i,\, s_i$ were obtained in the same way as in Example 4.10.

\bigskip

\centerline{{\bf Table 2}}

\medskip

\hspace{2mm}
\begin{tabular}{|*{11}{p{25pt}|}}\hline \rule{0pt}{15pt}
~~No. & ~~~n & $~~\mu$ & $~~k_1$ & $~~k_2$ & $~~s_1$ & $~~s_2$ & $~~r_1$ & $~~r_2$ & $~~l_1$ & $~~l_2$\\ \hline

$~~~2$ & $~~~0$ & $-\frac{7}{4}$ & $-\frac{25}{21}$ & $~~~\frac{6}{7}$ & $-\frac{10}{21}$ & $~~\frac{8}{7}$ & $-\frac{1}{2}$ & $~-\frac{6}{5}$ & $~~\frac{5}{4}$ & $~~\frac{9}{10}$ \\ \hline

$~~~4$ & $~~~0$ & $-1$ & $-\frac{1}{2}$ & $\sigma+1$ & $~~~\frac{3}{2}$ & $~~\sigma$ & $\frac{3}{3+4\sigma}$ & $-\frac{2 \sigma}{3+4\sigma}$
& $\frac{1}{3+4\sigma}$  & $\frac{2 (1+\sigma)}{3+4\sigma}$ \\ \hline

$~~~5$  & $~~~0$  & $-3 $ & $-\frac{25}{24}$ & $~~~\frac{3}{8}$ & $-\frac{5}{24}$ & $-\frac{1}{8}$ & $~~1$ & $~~-\frac{3}{5}$ & $-5$ & $-\frac{9}{5}$ \\ \hline

$~~~6$ &  $~~~1$ & $-\frac{7}{5}$ & $~~~\frac{2}{7}$ & $-\frac{1}{28}$ & $~~~\frac{5}{7}$ & $~~\frac{15}{28}$ & $-4$ & $~~~~~3$ & $~~\frac{8}{5}$ &   $~~\frac{1}{5}$ \\ \hline

\end{tabular}

\bigskip

\noindent  In the third row the constant $\sigma$ is an arbitrary parameter and therefore we found an explicit general solution of (\ref{mastereq}) with $\mu=-1,\, b=0$. The limit $\sigma\to \infty$ in this solution, together with a proper rescaling of $\tau$, yields the following particular solution \begin{equation}\label{eq2}
V\,=\,\tau^{-\frac{1}{2}} \, e^\tau,  \qquad e^{q_{2}}\,=\, \tau^{\frac{3}{2}}\, e^\tau.
\end{equation}

Given a rational solution of the Riccati equation (\ref{ric}), we can find a general solution $\tau(x)$ in the standard way.  On occasion the integral
for $q_2$ in (\ref{Sasha02Param}) can be found in elementary functions.

{\bf Example 4.11}. Consider the family 5 with $n=1.$ In this case $\mu\,=\,-\,2\,$ and
\begin{eqnarray} \nonumber
&&\tau(x)\,=\,\frac{c\, +\,x\, \sqrt{x^2\,-1}\,- \ln \left(\sqrt{x^2\,-1}\,+\,x\right)}{8 \left(c\,x\,+\,\sqrt{x^2\,-1}\,-\,x\, \ln \left(\sqrt{x^2\,-1}\,+\,x\right)\right)}\,,\\
&&e^{q_2(x)}\,=\,\frac{\left(c\, x\,+\,\sqrt{x^2\,-1}\,-\,x \,\ln \left(\sqrt{x^2\,-1}\,+\,x\right)
\right)^{3/4}}{\left(x^2\,-\,1\right)^{3/8}\,
(c\,+\,x\, \sqrt{x^2\,-1}\,-\,
\ln \left(\sqrt{x^2\,-\,1}\,+\,x\right))^{\frac{1}{2}}}\,,\nonumber\\
&&V(x)\,=\,\frac{\left(\,c\, x\,+\,\sqrt{x^2\,-\,1}\,-\,x \,\ln \left(\sqrt{x^2\,-\,1}\,+\,x\right)
\right)^{1/4}}{64 \,\left(x^2\,-\,1\right)^{9/8}\,(c\,+\,x\, \sqrt{x^2\,-1}-\,
\ln \left(\sqrt{x^2\,-\,1}\,+\,x\right))^{-\frac{1}{2}}\,}\nonumber,
\end{eqnarray}
where $c$ is an integration constant.

\medskip

We have already constructed two solutions of the master equation (\ref{mastereq}) with $b=0$ (see formulas (\ref{eq1}) and (\ref{eq2})) of the form
\begin{equation}\label{expparam}
V\,=\,\tau^{z_1}\,e^{z_2 \tau}, \qquad e^{q_{2}}\,=\, \tau^{z_3} \, e^{ \tau},
\end{equation}
where $z_1,\,z_2,\,z_3$ are constants.
A direct substitution of the ansatz (\ref{expparam}) into (\ref{mastereq}) with $b=0$ shows that there  exist only four solutions of such type. Thus besides the potentials (\ref{eq1}) at $\mu=0$ and (\ref{eq2}) at $\mu=-1$, there are two more solutions:
\begin{equation}\label{eq3}
 V \,=\, \tau^{3/4}\, e^{\tau}, \qquad e^{q_2} \,=\, \tau^{1/4}\, e^{\tau}\quad at  \quad \mu\,=\,0,
\end{equation}
\begin{equation}\label{eq4}
 V \,=\, \tau^{1/4} \,e^{\tau}, \qquad e^{q_2}\, =\, \tau^{3/4}\, e^{\tau}\quad at  \quad \mu\,=\,-2.
\end{equation}
Since involution (\ref{tr2}) and reparameterization $\tau\to - \tau$ map (\ref{expparam}) to a solution of (\ref{mastereq}) with $a=0$ of the same form (\ref{expparam}), where $z_3\to - z_3$ and $z_2\to -z_2,$ we have four more solutions corresponding to $a=0$. Thus altogether there are eight potentials of such type.

\begin{figure}[hbt]
\begin{center}
\iffigs
\includegraphics[height=55mm]{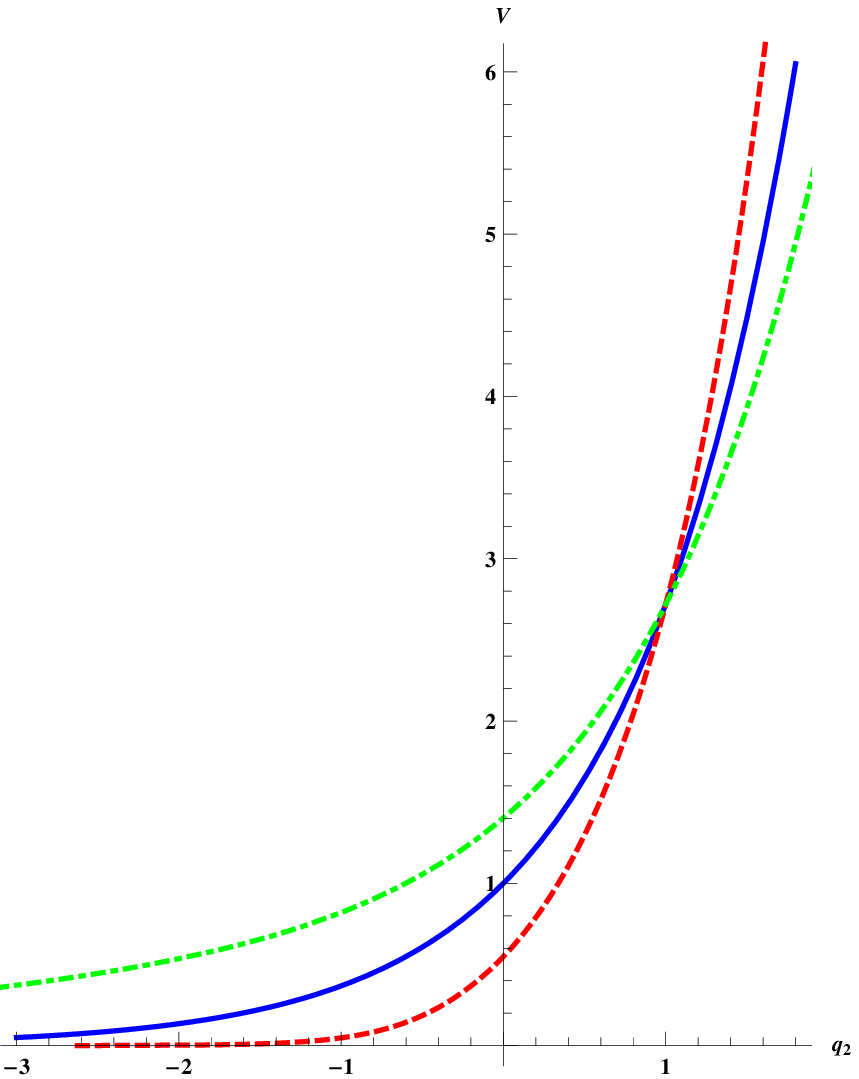}
\includegraphics[height=55mm]{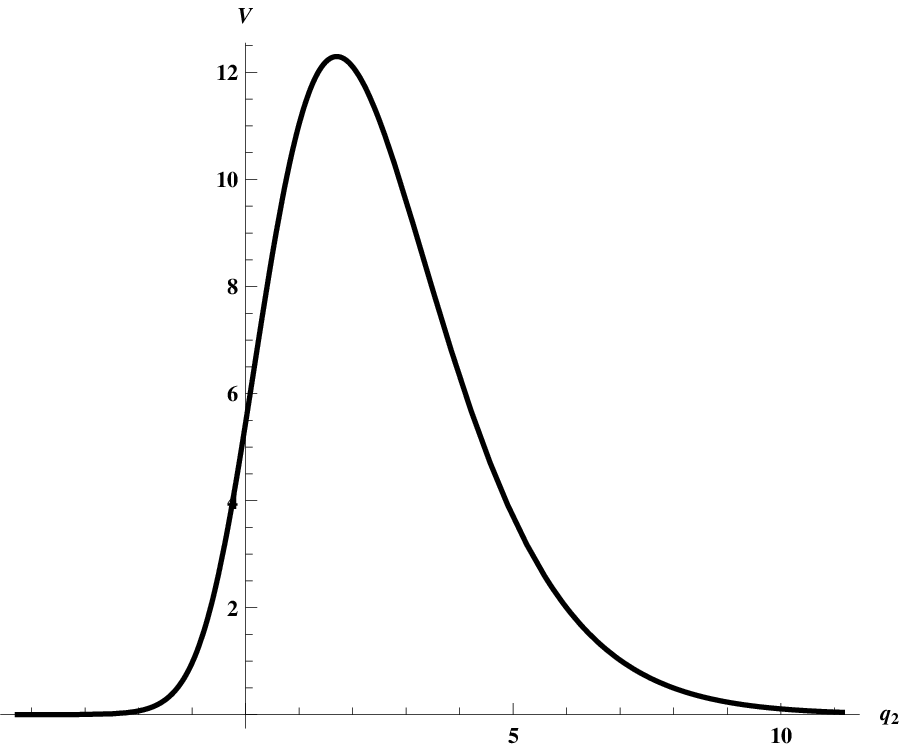}
\includegraphics[height=55mm]{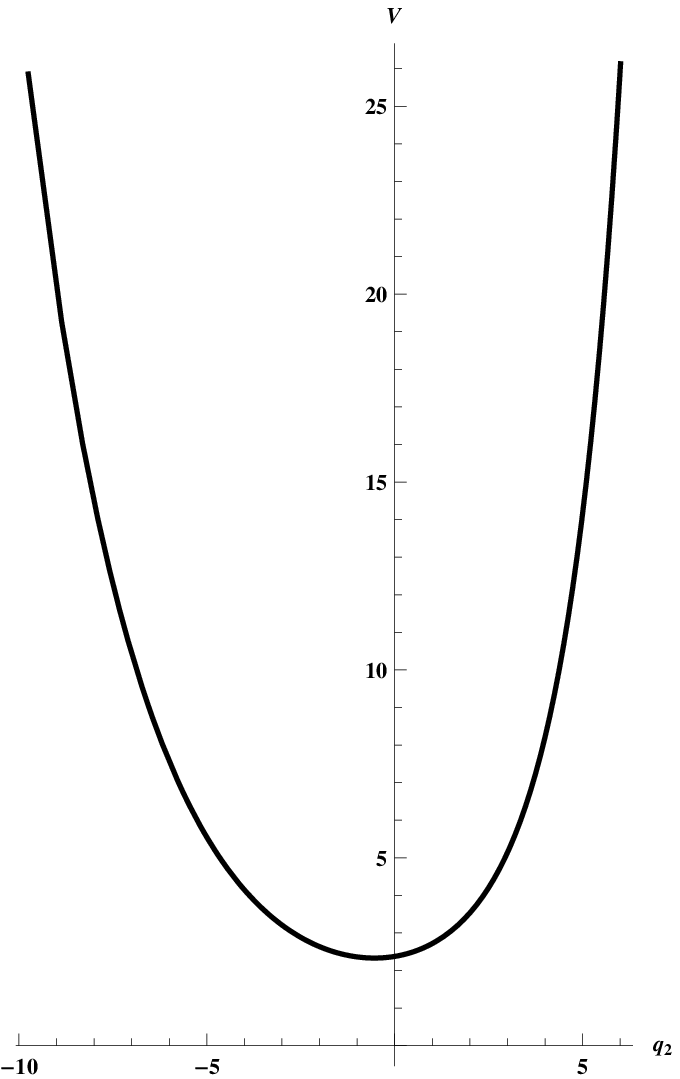}
\else
\end{center}
 \fi
\caption{\it
Left panel:
$\{V\,=\,t^{\frac{3}{4}}\,e^t,\,
q_2\,=\,t \,+\,\frac{1}{4}\ln(t)\}$   (dashed line),
$V\,=\,e^{q_2}$  (continuous line), and
$\{V\,=\,t^{\frac{1}{4}}\,e^{t},\,
q_2\,=\,t \,+\,\frac{3}{4}\ln(t)\}$   (dashed-dotted line).
Middle panel: $\{V\,=\,
30 \,t^{\frac{3}{2}}\,e^{-t},\,
q_2\,=\,t \,+\,\frac{1}{2}\ln(t)\}$.
Right panel:
$\{V\,=\,t^{-\frac{1}{2}}\,e^{t},\,
q_2\,=\,t \,+\,\frac{3}{2}\ln(t)\}$.
 }
\label{pot2}
 \iffigs
 \end{center}
  \fi
\end{figure}

It is easy to see that if $z_3 > 0,$ then for $\tau \in (0,\,+\infty)$ the function $q_2(\tau)$ ranges from $-\infty$ to $+\infty$ and we obtain a potential $V(q_2)$ defined on the whole real line of $q_2.$ We present plots of the four integrable potentials of the form (\ref{expparam}) with $z_3>0$ and the maximally superintegrable exponential potential (\ref{equ}) at the critical value $\lambda=1$ (Fig. \ref{pot2}).

\section{Conclusions and Outlook}

\subsection{Main results}

\qquad In the present paper we derive the master equation (\ref{mastereq}) for the potential $V.$ The function $V(q_2)$ satisfies the nonlinear ODE (\ref{mastereq}) if and only if the Hamiltonian $H$ of the form (\ref{HHour}) possesses a cubic integral  at the level $H=0 $. The master equation contains the constants  $\mu, a$, and $b$, where $\mu$ is an essential parameter, whereas $a$ and $b$ can be reduced (if $\mu\ne 0$) by admissible transformations (\ref{tr1}), (\ref{tr2}) to one of the following three alternatives $a=b=1,$ $a=-b=1,$ or  $b=0, \, a=1$. Notice that in the case $\mu=0$ the ratio $\frac{a}{b}$ becomes an essential continuous parameter.

In the case  $b=0, \, a=1$  for arbitrary $\mu$  we constructed general solution (\ref{gensol}) of master equation (\ref{mastereq}) in terms of the associated Legendre functions. We described the values of $\mu$ such that the general solution can be expressed in terms of elementary functions. These values are splitted into six families. We constructed new integrable potentials corresponding to simplest representatives of these families.

We also derived the master equations that provide the existence of additional linear and quadratic in momenta integrals and presented the complete list of all their  solutions. We proved that there are no cases with linear and quadratic integrable  different from the ones discussed in \cite{FSS,FS2}. We verified that both sporadic and non-sporadic integrable potentials from \cite{FSS,FS2} belong to this list. Furthermore, we demonstrated that all the sporadic potentials are actually very particular cases of the non-sporadic ones.

\subsection{Background and perspectives}

{}~

{\bf 1. Towards general solution of master equation with $a\, b\neq 0$}

{}~

In the future we plan to investigate master equation (\ref{mastereq}) with $a b\ne 0$. By now we have found several its solutions for special values of the parameter $\mu$   (see subsection 4.1.1). We believe that for generic $\mu, \,a$ and $b$ the general solution of (\ref{mastereq}) can be represented in a parametric form in terms of special functions.

{}~

{\bf 2. Analytic properties of solutions}

{}~

\begin{figure}[hbt]
\begin{center}
\iffigs
\includegraphics[height=70mm]{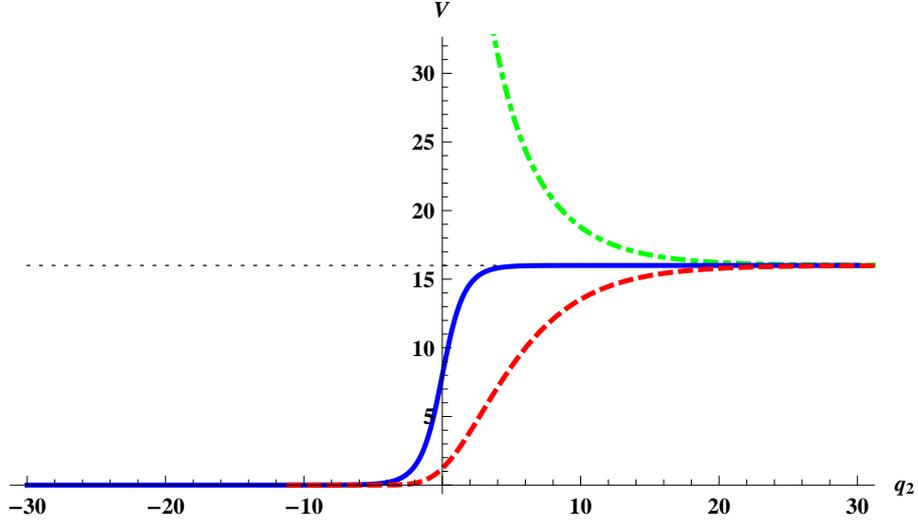}
\else
\end{center}
 \fi
\caption{\it The figure displays the three integrable positive-definite potentials: 1) $\{V\,=\,16\,t^2,\,\, q_2\,=\,\ln(t) \,-\,4\ln(1\,-\,t), \, 0<t<1\}$ (dashed line), 2) $V\,=\,\frac{32}{\pi}\,\arctan(e^{q_2})$ (continuous line), and 3) $\{V\,=\,16\,t^2,\, q_2\,=\,\ln(t) \,-\,4\ln(t\,-\,1), \, 1<t< +\infty\}$ (dashed-dotted line). The second potential is the middle solution at $\{C_1\,=\,0,\, C_2\,=\,\frac{32}{\pi}\}$ in eq. (\ref{sol5}). The first and the third potentials, defined in the whole real-line range $-\infty < q_2 < +\infty$, are positive definite roots of quasi-polynomial equation (\ref{eqVAnyMu}) at $\{\mu\,=\,-\frac{1}{2},\, c_1\,=\,\mp\,\frac{1}{\sqrt{2}} ,\, c_2\,=\,-\frac{1}{4} ,\, c_3\,=\,1 \}$. The latter equation can  be explicitly resolved for $q_2$: $q_2\,=\,-4\ln\Big(\sqrt{2}\,V^{-\frac{1}{8}}\,-\,
\frac{1}{2\sqrt{2}}\,V^{\frac{3}{8}}\Big),$ where $V \in (0,\,16)$ and $q_2\,=\,-4\ln\Big(-\sqrt{2}\,V^{-\frac{1}{8}}\,+\,
\frac{1}{2\sqrt{2}}\,V^{\frac{3}{8}}\Big),$ where $V \in (16,\,+\infty)$, respectively.}
\label{pot1}
 \iffigs
 \end{center}
  \fi
\end{figure}

We also plan to investigate analytical properties of the potentials obtained in Section 4. For instance, it follows from the explicit form of the coefficient $A_{13}$ in (\ref{mastereq}) that solutions with asymptotics $V(q_2)\to const \ne 0$ as $q_2 \to \infty$ should exist for $\mu=-\frac{1}{2},$ $\mu=-\frac{3}{2},$ and $\mu=-1$. Three examples of such solutions, which might be demanded in the developing of cosmological inflationary scenarios, are presented below (see Fig. \ref{pot1}).
The new potential (dashed line) looks qualitatively similar to the potential examined in \cite{FSS} (continuous line): they both never vanish and describe essentially finite (deformed) potential  steps. There is one more potential of the same type plotted by the dashed-dotted line (Fig. \ref{pot1}). This potential is given by eq. (\ref{param}) with $\{\mu\,=\,-1, \, \sigma\,=\,-1\}$ from Table 2. The detailed analysis of this subject is out of the scope of the present paper and will be discussed elsewhere.

{}~

{\bf 3. Superintegrable cosmological models}

{}~

It would be interesting to find all superintegrable Hamiltonians $H$ of the form (\ref{HHour}), i.e. Hamiltonians possessing two additional integrals of motion $h_1$ and $h_2$, which, together with $H,$ are functionally independent at the level $H=0$. The phase space of the corresponding dynamical systems is four-dimensional and their trajectories are defined by the levels $h_1=c_1,\, h_2=c_2, \, H=0$ of the integrals. In other words, any superintegrable model is the maximally superintegrable one. Besides already discussed potentials (\ref{equ}) we would like to present a few more examples of superintegrable Hamiltonians:

 {\bf Example 5.1}.  The Hamiltonian (\ref{HHour}) with
 $$
 V(q_2)=c_1\, e^{-\frac{q_2}{5}}+c_2\, e^{\frac{3 q_2}{5}}
 $$
 possesses two integrals of second degrees:
 $$
  h_1=2\, e^{\frac{-4 q_1-4 q_2}{5}}\,p_2^2 +  2\,  e^{\frac{-4 q_1-4 q_2}{5}}\,p_1 p_2 - c_1\, e^{\frac{q_1-5 q_2}{5}}+3 c_2\, e^{\frac{q_1-q_2}{5}},
 $$
 and
 $$
 h_2=\Big(6 c_1\, e^{\frac{-2 q_1-2 q_2}{5}} +18 c_2  e^{\frac{-2 q_1+2 q_2}{5}}  \Big)  \, p_2^2+ \Big(6 c_1\, e^{\frac{-2 q_1-2 q_2}{5}} -18 c_2  e^{\frac{-2 q_1+2 q_2}{5}}  \Big) \, p_1 p_2$$
 $$ -c_1^2\, e^{\frac{3 q_1-3 q_2}{5}}+  6 c_1 c_2\, e^{\frac{3 q_1+q_2}{5}} - 9 c_2^2\, e^{\frac{3 q_1+5 q_2}{5}}.
 $$

 {\bf Example 5.2}.  The  Hamiltonian with the potential
$$
V(q_2)\,=\,c_1\, e^{-\frac{q_2}{3}}\,+\,c_2\, e^{\frac{q_2}{3}}
$$
has integrals of first and second degrees
$$
h_1\,=\,\Big(c_1\, e^{\frac{-q_1\,+\,q_2}{3}}\,-\,c_2\,  e^{\frac{-q_1\,-\,q_2}{3}}\Big)\, p_1\,-\,\Big(c_1\, e^{\frac{-q_1\,+\,q_2}{3}}\,+\,c_2\,  e^{\frac{-q_1\,-\,q_2}{3}}\Big)\, p_2,
$$
$$
h_2\,=\,2 e^{\frac{-2 q_1\,+\,2 q_2}{3}}\,p_2^2\,-\,2 e^{\frac{-2 q_1\,+\,2 q_2}{3}}\,p_1 p_2\,-\,c_1\,  e^{\frac{ q_1\,+\,3 q_2}{3}}\,+\,c_2\, e^{\frac{ q_1\,+\, q_2}{3}}
$$
with the Poisson relation
$$
\{h_1,\, h_2\}\,=\,-\frac{4 c_2^2}{3}.
$$

{\bf Example 5.3}.  In the case
$$
V(q_2)\,=\,c_1\, e^{\frac{ q_2}{2}}\,+\,c_2\, e^{2 q_2}
$$
there exist the integrals of second and third degrees
$$h_1\,=\,2 e^{\frac{-3 q_1\,-\,3 q_2}{2}}\,p_2^2\,+\,2 e^{\frac{-3 q_1\,-\,3\, q_2}{2}}\,p_1\, p_2\,-\,c_1\,e^{\frac{- q_1\,-\,2 q_2}{2}}\,-\,4\, c_2\,e^{\frac{-q_1\,+\, q_2}{2}},$$
$$ h_2\,=\,2 e^{\frac{-3 q_1\,-\,3 q_2}{2}}\,p_2^3\,+\,2\, e^{\frac{-3 q_1\,-\,3 q_2}{2}}\,p_1\, p_2^2\,+\,\Big(c_1\,e^{\frac{- q_1\,-\,2 q_2}{2}}\,-\,8 \,c_2\,e^{\frac{-q_1\,+\, q_2}{2}}\Big)\, p_1.$$
They satisfy the following quadratic Poisson relation:
$$
\{h_1,\, h_2\}=\frac{3}{2}\,h_1^2, \qquad \mbox{mod}\,H=0.
$$

{}~

{\bf 4. Geometry of integrable cosmological models}

{}~

Alternative approaches to classification of integrable cosmological potentials are related to properties of the metrics \begin{equation}\label{metric}
d\,s^2=e^{q_1} V(q_2) \Big(d q_1^2-d q_2^2\Big)
\end{equation}
associated with Hamiltonains (\ref{HHour}) at the level $H=0$.
For example, it is well known  that any Killing tensor of the rank $k$ for (\ref{metric}) generates an additional integral of degree $k$ in momenta.
Killing tensors in the cases $k=2$ and $k=3$ were considered in  \cite{Rosquist:1991} and  \cite{Karlovini:1998kc}, respectively.
In the case $k=2$ the authors found all potentials from Section 3.1 except for (\ref{sol2}), (\ref{sol7}) and for first two solutions in (\ref{sol5}). For the case $k=3$  the analog of master equation (\ref{mastereq}) for  the quasi-K\"ahler potential $E=(\partial^2_ {q_1}-\partial^2_{q_2}) \, V$ was presented in \cite{Karlovini:1998kc}.  However no any new integrable potentials $E$ or $V$ were found there.

Global properties of metric (\ref{metric}) for some integrable potentials $V$ were investigated in \cite{Valent}. It would be interesting to examine the metrics corresponding to new integrable potentials constructed in the present paper.

{}~

{\bf 5. Embedding of integrable cosmological potentials into supergravities}

{}~

One more interesting direction is to explore different inclusions of integrable potentials found in the present paper into various supergravities. Embedding of the integrable potentials constructed in \cite{FSS,FS2} into $N=1$ supergravity by means of superpotential (F-term) is quite difficult task and was realized only in a very few cases \cite{FST1}. Although quite recent interesting paper \cite{FR} opens new perspectives in this direction. Every positive definite potential can be  minimally included as a D-term \cite{FKLP}, which was studied in \cite{FS2,FS1,FFS} for some of integrable potentials \cite{FSS}.

It would be also interesting to construct and analyze cosmological models admitting a quartic first integral and we hope to address to this task in the future.

\bigskip

\bigskip

{\bf Acknowledgments}

We thank P. Fr\'e, S. O. Krivonos, D. Sternheimer and A. I. Zobnin for useful discussions.
V. S. thanks JINR (Dubna) for the hospitality extended during his visits.
V. S. was supported by RFBI-grant No. 16-01-00289.
The work of A. S. was partially supported by RFBR grants
No. 15-52-05022 Arm-a and No. 16-52-12012-NNIO-a, DFG grant
Le-838/12-2 and by the Heisenberg-Landau program.


\end{document}